# Gravity of Accelerations on Quantum Scales and its Consequences

C Sivaram
Indian Institute of Astrophysics
Bangalore - 560 034, India
e-mail: sivaram@iiap.res.in

Kenath Arun
Christ Junior College
Bangalore - 560 029, India
e-mail: kenath.arun@cjc.christcollege.edu

Kiren O V
Christ Junior College
Bangalore - 560 029, India
e-mail: kiren.ov@cjc.christcollege.edu

B N Sreenath
S N Bhat P U College of Science & Commerce
Bangalore – 560050, India
e-mail: bnsreenath@yahoo.co.in

Abstract

Gravity stands apart from other fundamental interactions in that it is locally equivalent to an accelerated frame and can be transformed away. Again it is indistinguishable from the geometry of space-time (which is an arena for all other basic interactions), its strength being linked with the curvature. This is a major reason why it has so far not been amenable to quantisation like other interactions. It is also evident that new ideas are required to resolve several conundrums in areas like cosmology, black hole physics, and particles at high energies. That gravity can have strong coupling at microscales has also been suggested in several contexts earlier. Here we develop some of these ideas, especially in connection with the high accelerations experienced by particles at microscales, which would be interpreted as strong local gravitational fields. The consequences are developed for various situations and possible experimental manifestations are discussed.

# 1 Introduction

Gravity is one of the four fundamental interactions underlying the plethora of physical phenomena describing forces between particles and fields in space time. It is distinct from other forces like electromagnetism or nuclear in that it is locally equivalent to an accelerated frame and is described by the curvature of space time. Moreover it couples universally to all other fields. So gravity stands apart from other interactions and its unification with other forces is yet to be accomplished. A basic tenet of general relativity is the equivalence principle, i.e. a gravitational field is locally equivalent to an accelerated frame. Lorentz invariance corresponds to Inertial frames (IF's), moving with uniform velocity or at rest. They are associated with absence of forces (or accelerations) and are described in flat, Minkowskian space time (no gravitational field). Inertial frames are an idealization.

## 2 Acceleration at microscales and strong gravity

In practice subatomic particles feel forces such as electromagnetic or strong interactions over quantum scales of length and time. A particle initially at rest could be accelerated to high velocities over short time scale, i.e. over $\sim 10^{-16} - 10^{-23} s$. Typically this acceleration occurs over a time scale of $t \sim \hbar/mc^2$ (for a particle of mass m), so that acceleration is $\sim \dfrac{c}{\hbar/mc^2}$, i.e.

$$a \sim \frac{mc^3}{\hbar} \tag{1}$$

For $m \sim 10^{-27} kg$ (proton mass), this corresponds to $a \sim 10^{32} m/s^2$.

So by equivalence principle this should correspond to a 'strong gravitational field', over a distance scale $\sim \hbar/mc$. For two protons, with this strong force (accelerating them) causing this acceleration, $a$, their equivalent mutual gravitational interaction is given by:

$$a \sim \frac{mc^3}{\hbar} \approx \frac{G_{eff} m}{r^2} \tag{2}$$

This implies on 'effective gravitational constant' of

$$G_{eff} \approx \frac{\hbar c}{m^2} \approx 10^{38} G_{Newton} \tag{3}$$

over a distance scale of $\sim \hbar/mc$.

Once the protons leave the interaction (acceleration) region they would have uniform velocity ~c, so that once again we have IF and Lorentz invariance. Thus the time scale for presence of a non IF (acceleration) is $\hbar/mc^2 \approx 10^{-23} s$.

So this high acceleration is equivalent to a strong gravitational field with an effective $G \sim 10^{38} G_{Newton}$ over a distance scale $\sim \hbar/mc$.

Indeed this picture provides an additional framework for the strong gravity hypothesis (in considering particle interactions in the micro world) proposed several years ago.

For a detailed description of several interesting consequences of the approach see the extensive review article [1, 2] and references therein and more recently the chromo-gravity approach [3, 4, 5]. The past literature on this is extensive [6, 7, 8, 9, 10].

In this picture the short range hadronic gravitational interactions are mediated by massive spin-2 mesons (same quantum numbers as graviton). Starting with linear spin-2 field equation and successively adding all non-linear self interactions (of a non-abelian field line gravity) one does get Einstein type field equations with large curvature and strong coupling.

More recently [11] we have the ADD conjecture that gravity becomes strong at electroweak scale, giving an extra dimension of few microns. Gravity weakens much faster in higher dimensions. Hence we have an exponential short range gravitational field with large coupling (with large acceleration at the quark scale being the underlying feature).

Indeed from the Riemann equation:
$$a^\mu = R^\mu_{\alpha\beta\gamma} u^\alpha u^\beta n^\gamma \quad (4)$$

this large acceleration is also equivalent to a high curvature (over a subatomic scale), i.e.
$$R \sim \left(\frac{mc}{\hbar}\right)^2 \approx 10^{24} m^{-2} \quad (5)$$

A bunch of such particles in the high acceleration region (when they are being accelerated over quantum distance and time scales) would also experience a temperature
$$T \sim \frac{\hbar a}{2\pi k_B c} \quad (6)$$

This implies
$$T \sim \frac{mc^2}{2\pi k_B} \approx 2 \times 10^{12} K \quad (7)$$

Indeed in high energy collisions, there is a transverse thermal distribution of particles with this temperature, corresponding to the so called Hagedorn temperature. [1, 2]

## 3 Effective gravitational constant and modification of General Relativity

As the upper limit to particle mass m is the Planck mass $m_{Pl} \sim \left(\frac{\hbar c}{G}\right)^{1/2} \approx 10^{19} GeV$ this implies a maximum acceleration of: [12]

$$a_{max} \sim \frac{m_{Pl} c^3}{\hbar} \approx 10^{51} m/s^2 \qquad (8)$$

Corresponding to a highest curvature $\approx c^3/\hbar G \approx 10^{61} m^{-2}$ with a maximum temperature:

$$T_{max} \sim \frac{m_{Pl} c^2}{2\pi k_B} \approx 10^{32} K \qquad (9)$$

These upper limits (arising from quantum constraints on space time) could ameliorate classical singularities where, as $T \to \infty$, $R \to \infty$ as in the big bang singularity or in black hole collapse to a singular state [24, 25]. The classical GR could undergo modifications due to these limitations.

In this connection it is interesting that the use of minimum velocity [13], $\bar{c} \approx 3 \times 10^{-11} cm/s$, gives $a \approx 10^{-8} cm/s^2$, the MOND value for acceleration, when used in equation (1).

In another context we had suggested a Born – Infeld type of modification of GR, with a maximum curvature $R_{max}$, with a modified action like $\left(\frac{R}{1 - R/R_{max}}\right)\sqrt{-g}$, reducing to usual GR for $R << R_{max} \approx L_{Pl}^{-2}$. [14, 15]

In the above examples involving equation (2) and (3), we had an effective gravitational constant $G_{eff}$. Following a Sakharov type picture, we can write this as: [15, 16]

$$G_{eff} \approx \frac{c^5}{\hbar \int_0^{\omega_c} \omega d\omega} \qquad (10)$$

(which for $\omega_c \approx \frac{mc^2}{\hbar}$, gives $G_{eff} \sim \frac{\hbar c}{m^2}$ as above).

For the highest energy interactions, $\omega_c \approx \frac{m_{Pl} c^2}{\hbar}$ and this gives $G = G_{eff}$!

The action can be expanded in powers of the curvature, with a $R_{max} \approx \left(\frac{mc}{\hbar}\right)^2$. [16]

Essentially the effective gravity constant arises from all the zero point fluctuations of the various quantum fields (in curved space), which for an inertial frame has energy density:

$$\rho(\omega)d\omega \approx \frac{\omega^2}{\pi^2 c^3} \frac{\hbar\omega}{2} d\omega \qquad (11)$$

and for an accelerated frame this gets modified to:

$$\rho(\omega)d\omega \approx \frac{\omega^2}{\pi^2 c^3}\left(1+\frac{a^2}{c^2\omega^2}\right)\left[\frac{\hbar\omega}{2} + \frac{\hbar\omega}{e^{2\pi\omega c/a} - 1}\right]d\omega \qquad (12)$$

This would give rise to a temperature dependent $G_{eff}$ modified as:

$$G_{eff}(T) = G_{eff}\left(1 - \frac{T^2}{T_{max}^2}\right) \qquad (13)$$

(Vanishing for $T = T_{max}$, perhaps averting space-time singularities) [25]. Applications to the early universe and black hole collapse are relevant.

### 4 Energy levels in a gravitational field and solution of Dirac equation

As the high acceleration (at quantum scales) involve $\hbar$, we would have energy levels

$$E = \frac{n\hbar a}{c} \qquad (14)$$

Recent experiments have tried to look for these quantum levels, of neutrons in the earth's gravitational field [17, 18]. These levels could also be observable in high energy collisions.

The equation (14) can also be written as $E = ka$; $(\hbar = kc)$. [19]

These energy levels are given by using the Bohr-Wilson rule:

$$2\int_0^{E/mg} dy\left[2m(E - mgy)^{1/2}\right] = n\hbar \qquad (15)$$

giving the energy levels:

$$E_n = \left[\frac{9mgn^2\hbar^2}{32}\right]^{1/3} \qquad (16)$$

where *n* is an integer. (Varying accelerations can also be incorporated like $g = g(y)$, etc.

A solution of Dirac equation for electron energy levels in a gravitational field gives similar results, i.e. [2, 21]

$$[c\alpha.p + \beta(mc^2 + mgx)]\Psi(x) = 0 \qquad (17)$$

Using the two component form, we have:

$$\left[\left(1-\alpha\frac{\hbar g}{c}\right)p^2 + \frac{g^2}{c^2}\left(x+\frac{mc^2}{g}\right)^2\right]\phi_1 = \left(E^2 + \frac{\hbar g}{c}\right)\phi_1 \qquad (18)$$

(Similarly for $\phi_2$)

This gives the relativistic energy levels as:

$$E_n^2 = [2n\hbar gc + n^2\alpha g^2 c^2]^{1/2} \qquad (19)$$

For the case of a black hole we have:

$$a = \frac{c^4}{GM} \qquad (20)$$

where $M$ is the black hole mass. It turns out that there is a universal upper limit to Force as $F_{max} = \frac{c^4}{G}$.

This gives the corresponding Unruh temperature as:

$$T_u = \frac{n\hbar c^3}{GMk_B} \qquad (21)$$

So we have a prediction: we should get a set of energy levels apart from $n=1$ (Hawking temperature). This could imply gravitational Landau levels, $\hbar g/c$, playing role of gyrofrequency.

For primordial black holes with cut-off lifetime $t \sim 10^{18}s$, these levels would be $\sim 100 MeV$ (γ rays)!

## 5 Spiral trajectories of accelerated particles

The trajectories followed by accelerated particles (which lose energy by radiation and consequently have damped motion are well known to be exponential spirals. We have the equation of motion:

$$m\ddot{r} = \frac{e^2}{6\pi c^3}\dddot{r} + eV_{fluc} \qquad (22)$$

Essentially, $\ddot{r} \propto \dddot{r}$. This implies a path given by, $a = e^{kt}$, where $k$ is a constant. So the path is given by, $S = k''e^{k't}$; i.e. exponentially increasing (a logarithmic spiral).

This provides naturally an alternative picture for the inflation in the early universe. Thus we could have a primordial field accelerating the charged particles, where motion would be damped by radiation. So their path exponentially increases.

The high energy radiation could pair produce and create a shower of new particles. When the field weakens over a distance the particles stop being accelerated and motion becomes parabolic (or linear). These aspects will be quantitatively dealt with in a later paper. Again charged particles approaching the singularity in gravitational collapse could also be expected to follow such paths as their motion is damped by radiation emitted (as they are accelerated). The importance of spiral paths and the role of acceleration have been emphasized in ref. [19].

We recall the key property of the exponential spiral (logarithmic spiral) $R = R_0 e^{k\phi}$, i.e. the arc length from any point on the spiral to the centre (pole) is **finite** although it takes infinitely many rotations to reach the pole. This has implications for the singularity in black hole collapse. Again as Newton pointed out in his '*Principia*' if the law of gravity had been inverse cubic (this would be the case if there were an extra space dimension) an allowed orbit would be a logarithmic spiral (or exponential spiral).

Again relative acceleration between particles can be related to Riemann curvature of the underlying space time. Thus changes in acceleration would manifest as curvature changes. The differential acceleration would correspond to a temperature difference:

$$\Delta T = \frac{\hbar \Delta a}{2\pi k_B c} \qquad (23)$$

The $\Delta T$ can be related to curvature change $\Delta k$. So analogous to the heat conduction equation which describes heat flow via:

$$\frac{\partial T}{\partial t} \propto k \nabla^2 T \qquad (24)$$

We would have curvature flow (Ricci flow) in the dynamic space time of quantum micro-world.

$T$ is a constant corresponds to thermal equilibrium (no heat flow), while $k$ is a constant implies 'geometric' equilibrium of the underlying space time manifold. [20]

## 6 Space-time defects

Again space time defects could explain quantization of action (origin of $\hbar$) and quantum of electric charge [21]. Thus the torsion associated with a space-time defect, denotes the closure failure of a loop, in a surface element.

$$\oint Q dA = n \left( \frac{\hbar G}{c^3} \right)^{1/2} \qquad (25)$$

Quantification of closure in the Planck units could give a clue to the origin of $\hbar$, which would now be topological. Oscillations of the defect could give rise to acceleration through second time derivative of torsion.

Similarly second time derivatives of the curvature would give rise to angular accelerations, i.e. $\frac{\partial^2 \Delta\theta}{\partial t^2} \propto RdA$ over a torsion would arise from spin density, i.e. would give rise to a gravitomagnetic force (or gravitoferromagnetic effect) [22, 23]. Thus:

$$Q = \frac{4\pi G \sigma}{c^3} \quad (26)$$

$\sigma$ is the spin density ($N\hbar/r^3$ in a region of extent $r$, or $\sigma = N\hbar/r^3$).

So $Q$ is analogous to a magnetic dipolar field, falling off as $1/r^3$ (as seen from equation (26)).

Hence $\oint QdA$ gives 'gravitomagnetic flux' over surface $dA$, analogous to $\hbar c/e$ for the magnetic flux. This in turn confirms the possibility of gravitational Landau levels, $\left(n+\frac{1}{2}\right)Q.S$, $\frac{Q.S}{\hbar}$ is the 'gravito-gyro frequency'!

This has implications for strong gravitational fields, near black holes which can give a repulsive force for spinning black holes. Strong repulsive forces have been involved [19] to describe ejection of particles in jets around black holes. Torsion effects (gravitomagnetic forces) could give rise to a finite radius for collapsing matter preventing a singular state, i.e. inside the horizon the minimum radius would be:

$$R \approx \left(\frac{GS^2}{c^4 m}\right)^{1/3} \quad (27)$$

$S$ is the total spin. For a solar mass this gives $\sim 10^5 \, cm$. [24, 25]
For a derivation of equation (27), see [26]

### 7 Trapping of particles due to high acceleration

As high acceleration is locally a strong gravitational field, the usual criterion for gravitational confinement (of particles and fields) i.e. $\frac{Gm^2}{r} \geq mc^2$, can be written as:

$$r \leq \frac{Gm}{c^2}$$
$$\frac{\hbar a}{c} \geq mc^2 \quad (28)$$
$$a \geq \frac{mc^3}{\hbar}$$

We can get the same by substituting $\hbar c/m^2$ for $G_{eff}$ in equation (2) and (3). So this can be written as:

$$\frac{G_{eff} m^2}{r} \geq mc^2 \tag{29}$$

*TeV* black holes would correspond to $G_{eff} \approx \frac{\hbar c}{m_w^2} \geq mc^2$, where $m_w$ is the weak boson mass.

A whole range of values of $G_{eff}$ is possible as a function of energy ($G' = 10^{14} G$). *TeV* 'black holes' were first discussed in ref. [27] in this context. See also ref. [21, 28]. Also in connection with strings see ref. [26, 29, 30].

One can also have gravitational trapping in vortices if frequency $\omega > c/\lambda^{1/2}$, $\lambda$ being the vortex scale and $c$ the velocity of light.

Typically the time scale is $\sim \frac{\lambda^2}{\omega \nu}$, $\nu$ is the kinetic viscocity. Such phenomenon could occur in relativistic fluids (e.g. Quark gluon plasma, high temperature electron-positron plasma etc.) and could be tested in future experiments. As stated above strong gravitomagnetic forces could lead to formation of jets in relativistic fluids (also in accretion discs in close vicinity of massive black holes). The field could collimate the jet over long distances. At relativistic velocities, gravitomagnetic forces could be comparable to static gravitational forces. There are interesting spin-torsion, spin-orbit and spin-spin effects in such systems (brief discussion in ref. [21] and references therein). A more quantitative study is being explored. Such forces could even prevent the merger of black holes on their close approach.

**Reference**
[1] C. Sivaram and K.P. Sinha, Strong gravity, black holes, and hadrons. Phys. Rev. D. 16 (1977), 1975-1978.
[2] C. Sivaram and K.P. Sinha, Strong spin-two interaction and general relativity. Phys. Reports 51 (1979), 111-187.
[3] A. Salam and C. Sivaram, Strong Gravity Approach to QCD and Confinement. Mod. Phys. Lett. A. 8 (1993), 321-326.
[4] Y. Ne'eman and D. Sijaki, Origins of Nuclear and Hadron Symmetries. *Symmetries in Science V*. Plenum Press, New York (1991).
[5] A. Salam and J. Strathdee, Remarks on High-Energy Stability and Renormalizability of Gravity Theory. Phys. Rev. D 18 (1978), 4480-4485.
[6] L. Motz, Gauge invariance and the structure of charged particles. Il Nuovo Cimento 26 (1962), 672-697.

[7] A. Salam, Impact of quantum gravity theory on particle physics, in Quantum Gravity, an Oxford Symposium, Clarendon Press, Oxford (1975).
[8] N. Rosen, General relativity with a background metric. Found. Phys. 10 (1980), 673-704.
[9] D.D. Ivanenko, in Astrofisica e Cosmologia, Gravitazione, Quanti e Relativit`a, Centenario di Einstein, edited by M. Pantaleo and F. de Finis, Giunti-Barbera, Florence (1978).
[10] E. Recami et al, Microuniverses and 'strong black holes': A Purely geometric approach to elementary particles. arXiv:gr-qc/9509005 (1995).
[11] N. Arkani-Hamed, S. Dimopoulos and G. Dvali, The hierarchy problem and new dimensions at a millimeter. Phys. Lett. B 429 (1998), 263-272.
[12] C. Sivaram and V. de Sabbata, Maximum acceleration and magnetic field in the early universe. Astrophys. Space Sci. 176 (1991), 145-148.
[13] B.N. Sreenath, K. Arun and C. Sivaram, Is there lower limit to velocity or velocity change? Astrophys. Space Sci. 345 (2013), 209-211.
[14] C. Sivaram, Some implications of quantum gravity and string theory for everyday physics. Curr. Sci. 79 (2000), 413-420.
[15] V. de Sabbata, C. Sivaram and D.X. Wang, A Born-Infeld Type of Modification of General Relativity with Maximal Curvature. Acta Astrophysica Sinica 14 (1994), 11-15.
[16] C. Sivaram, Fundamental interactions in the early universe. Int. J. Theor. Phys. 33 (1994), 2407-2413.
[17] V.V. Nesvizhevsky et al, Quantum states of neutrons in the Earth's gravitational field. Nature 415 (2002), 297-299.
[18] T.J. Bowles, Quantum effects of gravity. Nature 415 (2002), 267-268.
[19] B.N. Sreenath, Identifying quantum-gravity field with exponentially varying accelerated (or gravity) field. (unpublished, 2010)
[20] C. Sivaram and K. Arun, Extended Equivalence Principle: Implications for Gravity, Geometry and Thermodynamics. Hadronic J. 35 (2012), 653-659.
[21] V. de Sabbata and C. Sivaram, *Spin and Torsion in Gravitation*. World Scientific, Singapore, 1994.
[22] V. de Sabbata and C. Sivaram, Magnetic fields in the early universe. Il Nuovo Cimento 102 (1988), 107-112.
[23] V. de Sabbata and C. Sivaram, Strong spin-torsion interaction between spinning protons. Il Nuovo Cimento A 101 (1989), 273-283.
[24] C. Sivaram and K. Arun, Some enigmatic aspects of the early universe. Astrophys. Space Sci. 334 (2011), 225-230.
[25] C. Sivaram and K. Arun, Enigmatic aspects of entropy inside the black hole: what do falling comoving observers see? Astrophys. Space Sci. 337 (2012), 169-172.

[26] Sivaram C and De Sabbata, Torsion as the basis for string tension. Annalen der Physik 503 (1990), 419-421.

[27] C. Sivaram and K.P. Sinha, A Finite Neutrino Rest Mass from General Relativity. Current Science 43 (1974), 165-168.

[28] V. de Sabbata and C. Sivaram, Torsion and the cosmological constant problem. Astrophys. Space Sci. 165 (1990), 51-55.

[29] C. Sivaram, String tension and fundamental constants in the early universe. Astrophys. Space Sci. 167 (1990), 335-340.

[30] V. de Sabbata and C. Sivaram, Strong Gravity and Strings. Annalen der Physik 503 (1991), 362-364.